# Highly efficient compact acousto-optic modulator based on a hybrid-lattice hollow core fiber

Ricardo E. da Silva, Jonas H. Osório, David J. Webb, Frédéric Gérôme, Fetah Benabid, Marcos A. R. Franco and Cristiano M. B. Cordeiro

**We demonstrate the acousto-optic modulation of a hybrid-lattice hollow core fiber (HL-HCF) for the first time. For many years, optical fibers with reduced diameters have been the main solution to increase the interaction of acoustic and optical waves. However, the high drive voltages and large modulator components still employed drastically affect the efficiency and miniaturization of these devices. Here, we experimentally show that combining Kagomé and tubular lattices in HL-HCFs allows for enhancing the amplification of the acoustic waves and the modulation of the guided optical modes thus providing high modulation efficiency even when using a fiber with a 240 μm diameter. To the best of our knowledge, the measured HL-HCF's modulation efficiency (1.3 dB/V) is the highest compared to devices employing reduced diameter fibers. Additionally, we demonstrate a compact acousto-optic modulator with driver dimensions smaller than the HL-HCF diameter. Overall, our results show a promising alternative to solve the compromise of speed, efficiency, and compactness for integration with microscale all-fiber photonic devices.**

## I. INTRODUCTION

All-fiber acousto-optic modulators (AOMs) enable remarkable applications in electrically tunable wavelength filters, frequency shifters, heterodyne vibration sensors, and pulsed fiber lasers [1], [2], [3], [4]. In these devices, the spectral and power properties of the optical fibers are dynamically tuned by the frequency and amplitude of the electrical signal, providing low loss and monolithic integration with fiber optic components.

This work was supported by the grants 2022/10584-9, São Paulo Research Foundation (FAPESP), 305321/2023-4, 309989/2021-3, and 305024/2023-0, Conselho Nacional de Desenvolvimento Científico e Tecnológico (CNPq), and RED-00046-23, Minas Gerais Research Foundation (FAPEMIG).

R. E. da Silva and C. M. B. Cordeiro are with the Institute of Physics Gleb Wataghin, University of Campinas (UNICAMP), Campinas, 13083-859, Brazil (resilva@unicamp.br).

J. H. Osório is with the Department of Physics, Federal University of Lavras (UFLA), Lavras, 37200-900, Brazil.

M. A. R. Franco is with Institute for Advanced Studies (IEAv), São José dos Campos, 12228-001, Brazil.

D. J. Webb is with the Aston Institute of Photonic Technologies (AIPT), Aston University, Birmingham, B4 7ET, UK. R. E. da Silva is also a visiting fellow at AIPT, Aston University.

F. Gérôme and F. Benabid are with the GPPMM Group, XLIM Institute, UMR CNRS 7252, University of Limoges, Limoges, 87060, France.

AOMs based on flexural acoustic waves couple the power between the guided modes in the optical fiber. Previous studies employing distinct fiber types show that the fibers' mechanical and optical properties play an important role in the modulated transmission spectrum, defining the tuning wavelength range, notch bandwidth, modulation depth, and overall modulator efficiency. Thus, AOMs with reduced diameter etched fibers and long interaction lengths have been used to enhance the modulating efficiency [1], [2], [4], [5], [6]. In contrast, the high voltages still employed decrease the overall AOM efficiency, requiring power amplifiers. As alternatives, suspended core fibers [7] and hollow core fibers [8], [9], [10] achieve significantly high modulation depths and efficiencies with no need for tapering or etching techniques. The increased acoustic and optical interactions impact the size reduction of the AOM components, switching time, and consumed electric energy.

Recently, we have demonstrated that in solid silica fibers, the acoustic wave is mostly concentrated over the cladding, thus having a weak overlap with the optical modes in the fiber core [11]. Besides, we have shown that tubular-lattice hollow core fibers (TL-HCF) significantly increase the acousto-optic interaction consuming lower acoustic energy [10], [11] due to the silica reduction and the out-of-center position of the cladding tubes over the fiber cross section. Furthermore, TL-HCFs provide stronger spatial overlap of the coupled optical modes, hence contributing to increase the modulation depth of the spectral resonances.

Here, we experimentally demonstrate the acousto-optic modulation of a hybrid-lattice hollow core fiber (HL-HCF), which displays a cladding formed by a Kagomé structure and a ring of tubes defining its air core. Optically, these two claddings decrease the confinement losses [12]. Additionally, due to its microstructure characteristics, the HL-HCF further reduces the silica content in the fiber cross section compared to other HCFs, while keeping the properties of tubular lattices that are beneficial for acousto-optic modulation.

In this context, we show herein that combining Kagomé and tubular lattices in the HL-HCF allows for attaining a modulation efficiency of 1.3 dB/V, which is the highest among other devices, even when using fibers with considerable smaller diameters. Additionally, we report on a compact acousto-optic modulator with smaller dimensions than the HL-HCF itself. We hence understand that our results offer a promising solution for integrating high speed, efficiency, and compactness into microscale all-fiber photonic devices.

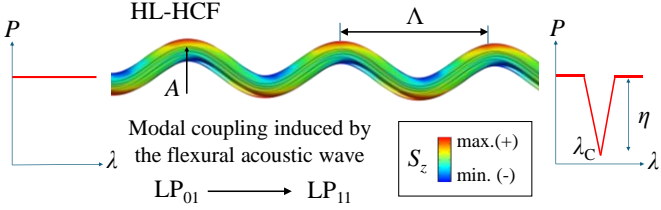

**Fig. 1**. Illustration of a hybrid-lattice hollow core fiber (HL-HCF) modulated by a flexural acoustic wave of amplitude $A$ and period $\Lambda$. The acoustic wave induces a longitudinal strain $S_z$, coupling power between the fundamental mode $LP_{01}$ and the higher-order mode $LP_{11}$, inducing a notch in the power ($P$) transmission spectrum at the resonant wavelength $\lambda_C$.

## II. ACOUSTO-OPTIC MODULATION OF HCFS

When optical fibers are modulated by flexural acoustic waves, the refractive index profile over their cross section is altered. In HCFs, this variation is mainly caused by the fiber's geometric deformation that changes the optical path of the guided modes [11]. Fig. 1 illustrates a flexural acoustic wave with amplitude $A$, period $\Lambda$, and frequency $f$. The acoustically induced longitudinal strain $S_z$ causes a net change in the refractive index, which is non-uniform over the fiber cross section.

The acoustic wave couples the power from the fundamental mode $LP_{01}$ to a higher-order mode $LP_{1m}$ at the resonance wavelength $\lambda_C$ when the optical beatlength $L_B$ matches with the acoustic period $\Lambda$, *i.e.*, $L_B = \Lambda$, where [2], [5], [8],

$$L_B = \frac{\lambda_C}{n_{01} - n_{1m}}, \quad (1)$$

$n_{01}$ and $n_{1m}$ are the effective refractive indices of the modes $LP_{01}$ and $LP_{1m}$, and,

$$\Lambda = \left(\frac{\pi D c_{ext}}{2f}\right)^{\frac{1}{2}}, \quad (2)$$

where $D$ is the fiber diameter and $c_{ext} = 5740$ m/s is the extensional acoustic velocity. The effective indices $n_{01}$ and $n_{1m}$ can be calculated analytically as reported in [12]. As we will see in the following, we will use the latter equations to investigate the optical couplings in our acousto-optic configuration. The results and corresponding analyses can be found in the next section of the manuscript.

## III. EXPERIMENTAL SETUP

Fig. 2(a) shows the cross section of the HL-HCF, which is composed of a hybrid cladding formed by Kagomé-tubular lattices. The thickness of the struts in the Kagomé structure is 800 nm. The fiber core (34 µm diameter) is defined by a ring of six tubes of 20 µm in diameter and 1.25 µm thickness. The fiber is made of pure silica and its outer diameter is 240 µm. The inset in Fig. 2(a) shows a connecting tube of about 50 nm thickness between the Kagomé structure and a core tube, providing appropriate spacing between the two claddings and preventing leakage of the fundamental mode. Details about the HL-HCF design and fabrication are described in [12].

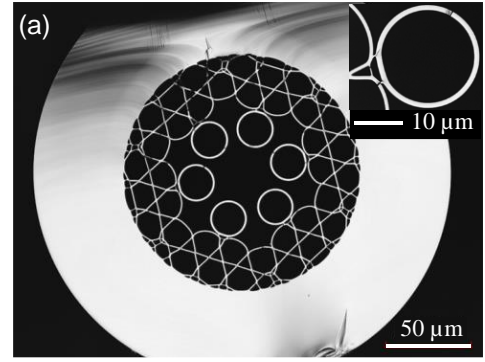

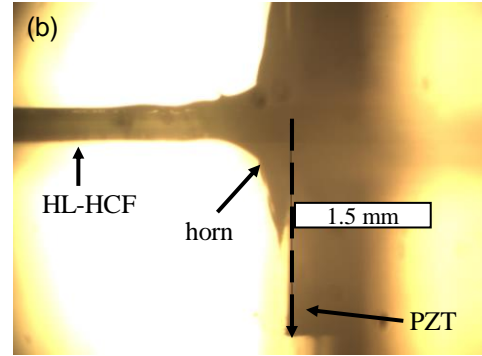

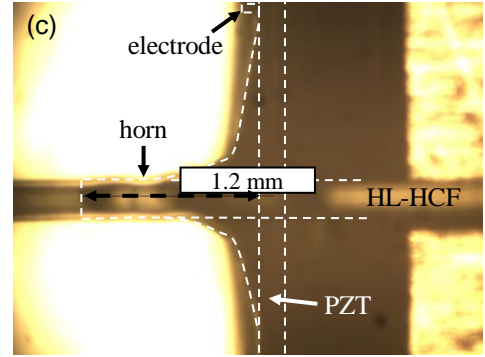

**Fig. 2**. (a) Cross section of the hybrid-lattice hollow core fiber (HL-HCF) showing in the inset a connecting tube between the Kagomé structure and a core tube. (b) Microscope view of the acousto-optic modulator (driver section) indicating the HL-HCF, acoustic horn and piezoelectric transducer (PZT). (c) Details of the acoustic horn indicating the electrode tip and the HL-HCF passing through the horn and PZT.

The acousto-optic modulator is composed of a HL-HCF segment, a piezoelectric transducer disc (PZT), and an axially connected acoustic horn, as shown in Fig. 2(b). The PZT is 200 µm thick and has a diameter of 3 mm. The horn is made by molding a fixing adhesive during its fluid state using the thin tip of a needle and the help of a microscope (MSC). Fig. 2(c) indicates the horn length of ~1.2 mm with a nearly exponential-like profile, decreasing from 2 mm to match the fiber diameter (240 µm). Note in Fig. 2(c) that the HL-HCF passes through the horn and PZT (a hole of about 500 µm is drilled in the PZT center). An electrode of 90 µm in diameter is connected to the PZT by means of a conductive silver adhesive. The other PZT side is connected to a thin metallic support (not visible in Fig. 2). The output fiber end is fixed forming a fiber interaction length of about $L \sim 7.7$ cm (the whole modulator including horn and PZT is 7.8 cm long).

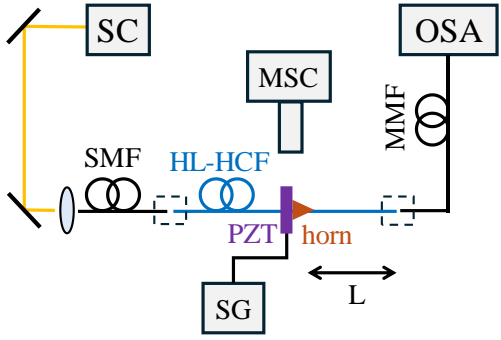

**Fig. 3**. Illustration of the experimental setup and acousto-optic modulator.

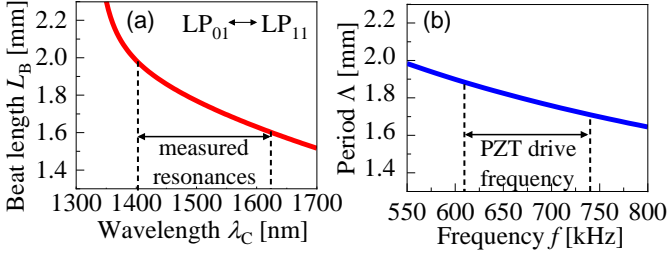

**Fig. 4**. (a) Analytical simulation of the optical beat length $L_B$ of the modes $LP_{01}$ and $LP_{11}$ in the HL-HCF and, (b) variation of the acoustic period $\Lambda$, indicating respectively the ranges of the experimentally measured resonances and drive frequency applied to the PZT.

Fig. 3 illustrates the experimental setup. A sinusoidal electrical signal with a maximum voltage of 10 V is applied to the PZT in the frequency range from $f$ = 611 to 733 kHz using a signal generator (SG).

The HL-HCF's spectrum is measured by using a supercontinuum source (SC) and an optical spectrum analyzer (OSA). The HL-HCF input is butt-coupled to a single mode fiber (SMF), previously aligned with the SC beam. The output is butt-coupled to a multimode fiber (MMF) connected to the OSA. The coupling alignments are performed using micrometer stages, mirrors, and a lens.

## IV. RESULTS AND DISCUSSION

Fig. 4(a) shows the simulated beat length $L_B$ of the modes $LP_{01}$ and $LP_{11}$ between $\lambda$ = 1300 - 1700 nm calculated using (1). Here, the effective refractive indices $n_{01}$ and $n_{11}$ were computed employing the model reported in [13] and the fiber parameters indicated in the last section. The acoustic period $\Lambda$ is calculated from $f$ = 550 to 800 kHz using (2).

We experimentally measured modulated resonances from $\lambda$ = 1401 to 1623 nm (222 nm range), indicating good agreement with the smooth central region of the curves in Fig. 4. Fig. 5 shows the measured modulated transmission spectrum of the HL-HCF, indicating the resonance with the highest achieved modulation depth (13.7 dB at $\lambda_{C2}$ = 1523 nm and $f$ = 636.3 kHz). The smaller notch at shorter wavelengths indicates acoustically induced birefringence of the mode $LP_{11}$ into the modes $LP_{11}^{Even}$ and $LP_{11}^{Odd}$. The average separation of the notches for the whole frequency range is 96 nm. This birefringence is caused by the non-uniform change of the refractive index over the fiber cross section. Having acoustically induced birefringence seems to be common in HCFs as seen in [9], [10] and explained in detail in [11]. The resonances' average half-height bandwidth is 20 nm.

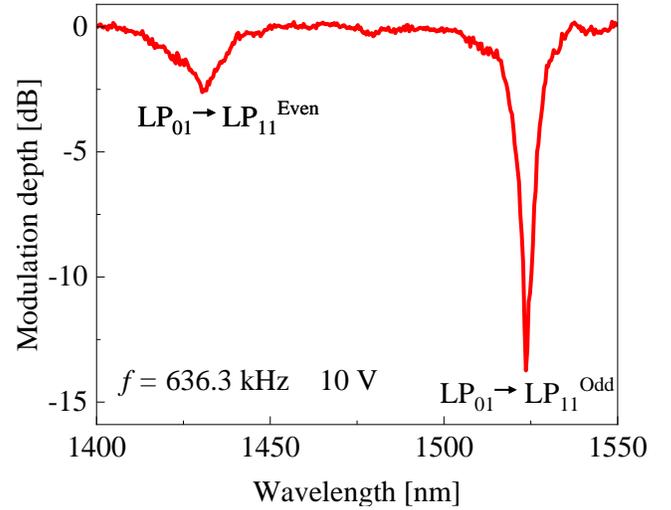

**Fig. 5**. Acoustically modulated resonance in the transmission spectrum of the HL-PCF.

The frequency tuning of the modes $LP_{11}^{Even}$ and $LP_{11}^{Odd}$ is shown in Fig. 6. The measured resonances cover tuning ranges of $\lambda$ = 1401 - 1521 nm ($LP_{11}^{Even}$ - black) and $\lambda$ = 1495 - 1623 nm ($LP_{11}^{Odd}$ - orange). The frequency tuning is almost linear for both modes ($r^2$ = 0.998). The spectral gap between $f$ = 650 and 733 kHz is caused by the intrinsic resonant frequency response of the PZT, which is commonly observed for most transducers employed in AOMs [4], [10]. Additionally, AOMs using coaxially aligned components might excite longitudinal acoustic waves, which cannot satisfy the phase-matching condition to couple the modes at the considered frequency range. Continuous tuning might be achieved by employing shear modes PZTs, which can be designed to oscillate transversally, favoring the excitation of flexural acoustic waves [9]. Overall, the HL-HCF provides continuous tuning for the entire effective frequency range, as seen in Fig. 4.

We compared the performance of our AOM with devices employing fibers with distinct diameters and geometries. The overall modulation efficiency (in dB/V) is estimated from the ratio of the maximum modulation depth and the respective consumed voltage. Table 1 summarizes the modulation efficiency, the fiber diameter, and the employed interaction length, which directly influence the overall modulation efficiency. The considered fibers include an etched SMF, an etched dispersion compensation fiber (DCF), a few-mode fiber (FMF), a photonic band gap (PBG-HCF), and a tubular-lattice (TL) HCF (references in Table 1). Note that the HL-HCF's modulation efficiency is 1.18x higher compared to the etched SMF (with ~60% longer length) and 2.6x higher than the etched DCF (with ~60% shorter length). Considering that the etched fibers are close in diameter, an average of ~2x higher efficiency is estimated for the HL-HCF, even employing a diameter 7x larger. Although the interaction length and the other modulator components can relevantly influence the overall efficiency, the dependence of the increasing efficiency with silica reduction is clearly noted for the distinct fibers in Table 1. As expected, the considered HCFs show decreasing silica content in the cross section, providing higher efficiencies compared to the solid fibers [11]. Overall, the HL-HCF structure with suspended tubes provides an extra degree of freedom to amplify the strain compared to the TL and PBG HCFs. The achieved modulation efficiency might be further improved by expanding the Kagomé cladding, reducing the thickness of the Kagomé structure and tubes, and increasing the air core. Further advances might also come from HL-HCFs with different dimensions and cladding designs. In particular, having fibers with larger diameters is significantly beneficial to reducing the switching time $\tau$ of AOMs. Considering the resonance parameters in Fig. 5, a switching time of $\tau = 33$ µs is estimated using $\tau = L/(2\sqrt{0.5\pi D c_{ext} f})$ [5], which is almost half compared to that corresponding to the etched DCF ($\tau = 57$ µs) employing similar frequency values [5].

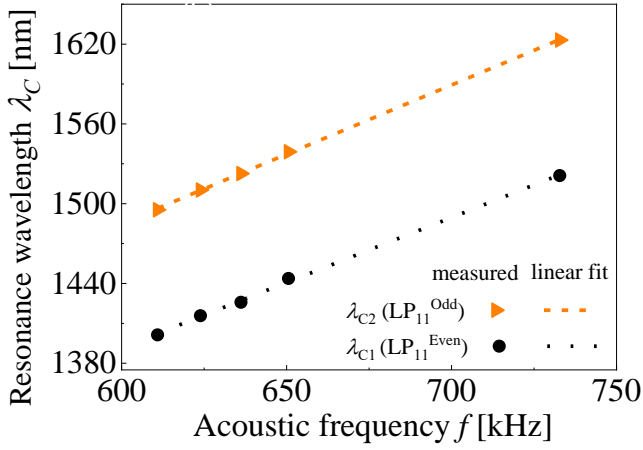

**Fig. 6**. Frequency tuning of the acoustically modulated resonances in the transmission spectrum of the HL-HCF.

## V. CONCLUSION

We have experimentally demonstrated the highly efficient interaction of a HL-HCF and flexural acoustic waves. A good agreement between the analytical simulations and measured resonances is observed over a spectral range of 222 nm (1401 to 1623 nm). A maximum modulation depth of 13.7 dB is achieved at 10 V. The AOM modulation efficiency of 1.3 dB/V is higher than in other devices that use reduced diameter fibers. This efficiency is remarkably 2x higher than a 35 µm etched fiber, indicating that the hybrid Kagomé-tubular lattice provides excellent amplification, hence increasing the induced strain in the fiber. The reduced acoustic horn (1.2 x 2 mm) and electrode (90 µm) contribute to relieving the load on the thin 200 µm PZT, enabling the acoustic modulation of a 240 µm fiber. The combined properties of the HL-HCF and modulator design provide the lowest consumed voltage (10 V), removing the need for power amplifiers. The short interaction length and large fiber diameter are promising to decrease the switching time of AOMs while providing better mechanical stability and protection. These features are highly attractive in applications requiring fast, compact, and efficient all-fiber integrated photonic devices.

TABLE I
ACOUSTO-OPTIC DEVICES BASED ON DISTINCT FIBER TYPES

| Fiber type | Diameter $D$ (µm) | Length (cm) | Efficiency (dB/V) | Reference |
|---|---|---|---|---|
| HL-HCF | 240 | 7.7 | 1.3 | This work |
| SMF – ET | 33 | 12.0 | 1.1 | [1] |
| PBG-HCF | 120 | 12.0 | 0.8 | [9] |
| DCF – ET | 35 | 4.8 | 0.5 | [6] |
| TL-HCF | 200 | 7.7 | 0.4 | [10] |
| FMF | 125 | N.A. | 0.2 | [2] |
| SMF | 125 | N.A. | 0.2 | [3] |

Geometric parameters and efficiencies of AOMs employing distinct fiber geometries (ET = etched). N.A.: not available.

ACKNOWLEDGMENT

We thank F. Delahaye and F. Amrani of the GPPMM Group, XLIM Institute, University of Limoges, for contributing to the fiber fabrication.